\documentclass[12pt]{article}
\usepackage{amssym}

\renewcommand{\theequation}{\arabic{section}.\arabic{equation}}

\def\R{\mbox{$\Bbb R$}}

\begin{document}
\baselineskip=20pt plus 1pt minus 1pt
\begin{center}

{\Large\bf\boldmath Disentangling $q$-Exponentials: A General Approach}

\bigskip

{\large\bf C.\ Quesne\footnote{\rm Physique Nucl\'eaire Th\'eorique et Physique
Math\'ematique, Universit\'e Libre de Bruxelles, Campus de la Plaine CP229, Boulevard du
Triomphe, B-1050 Brussels, Belgium; e-mail: cquesne@ulb.ac.be}}

\end{center}

\bigskip
\begin{abstract}
We revisit the $q$-deformed counterpart of the Zassenhaus formula, expressing the
Jackson $q$-exponential of the sum of two non-$q$-commuting operators as an (in
general) infinite product of $q$-exponential operators involving repeated
$q$-commutators of increasing order, $E_q(A+B) = E_{q^{\alpha_0}}(A)
E_{q^{\alpha_1}}(B) \prod_{i=2}^{\infty} E_{q^{\alpha_i}}(C_i)$. By systematically
transforming the $q$-exponentials into exponentials of series and using the conventional
Baker-Campbell-Hausdorff formula, we prove that one can make any choice for the bases
$q^{\alpha_i}$, $i=0$, 1, 2,~\ldots, of the $q$-exponentials in the infinite product. An
explicit calculation of the operators $C_i$ in the successive factors, carried out up to
sixth order, also shows that the simplest $q$-Zassenhaus formula is obtained for
$\alpha_0 = \alpha_1 = 1$, $\alpha_2 = 2$, and $\alpha_3 = 3$. This confirms and
reinforces a result of Sridhar and Jagannathan, based on fourth-order calculations.
\end{abstract}

\bigskip
\noindent
Running head: Disentangling $q$-Exponentials 

\newpage
%
%
\section{INTRODUCTION}

Disentangling the exponential of the sum of two noncommuting operators into an (in
general) infinite product of exponential operators involving repeated commutators of
increasing order is a problem that occurs in many fields of physics, such as statistical
mechanics, many-body theories, quantum optics and path-integration techniques (see,
e.g., Wilcox, 1967; Witschel, 1975; Suzuki, 1977; Hatano and Suzuki, 1991; Zhao, 1991;
Brif, 1996). In particular, such a procedure may be employed to provide some useful
approximation methods.\par
%
%
The problem is solved by applying the Zassenhaus formula, which was derived by Magnus
(1954) citing unpublished work by Zassenhaus. This formula is the dual of the
Baker-Campbell-Hausdorff (BCH) formula (Campbell, 1898; Baker, 1902, 1903, 1904a,
1904b; Hausdorff, 1906), expressing the product of two noncommuting exponential
operators as a single exponential operator in which the exponent is, in general, an infinite
series in terms of repeated commutators.\par
%
%
Since their advent (Drinfeld, 1987; Jimbo, 1985, 1986; Faddeev {\sl et al.}, 1988),
quantum groups and quantum algebras have had an ever-increasing and broader range of
applications in mathematics and physics (see, e.g., Majid, 1995; Chaichian and Demichev,
1996; Klimyk and Schm\" udgen, 1997). In developing noncommutative aspects of
$q$-analysis, there has been a growing interest in getting $q$-deformed counterparts of
standard results of conventional analysis, such as the BCH and Zassenhaus formulas.\par
%
%
In this respect, the simplest results are obtained for the Jackson $q$-exponential
$E_q(z)$ (Jackson, 1904), based on the use of the Heine basic numbers of base $q$,
$[n]_q \equiv (1-q^n)/(1-q)$ (Heine, 1847). This function is often referred to as the
maths-type $q$-exponential to distinguish it from the phys-type $q$-exponential for
which the symmetric $q$-numbers $[n]_q \equiv (q^n-q^{-n})/(q-q^{-1})$ are
employed.\par
%
%
A $q$-analogue of the BCH formula for $E_q(z)$ was derived by Katriel and Solomon
(1991). Later on, Katriel {\sl et al.} (1996) proposed a $q$-analogue of the Zassenhaus
formula, wherein the $q$-exponential of the sum of two non-$q$-commuting operators
is expressed as an (in general) infinite product of $q$-exponential operators involving
repeated $q$-commutators of increasing order. Recently, Sridhar and Jagannathan
(2002) derived another form of the $q$-Zassenhaus formula wherein, unlike in the Katriel
{\sl et al.} formula, the bases of the $q$-exponential factors in the infinite product are
not the same. In both works, the operators in the successive factors were determined up
to fourth order in the two operators.\par
%
%
Such results raise two questions: can one make any choice of bases for the
$q$-exponential factors in the $q$-Zassenhaus formula and, if so, for which choice of
bases does the formula take the simplest form? It is the purpose of the present paper to
answer both of these questions.\par
%
%
To be able to carry out the analysis in general terms without making any choice of bases
from the very beginning, we shall adopt another procedure than those previously
employed. It is based on the repeated use of the conventional BCH formula after
expressing every $q$-exponential as a standard exponential of a series (Hardy and
Littlewood, 1946).\par
%
%
This paper is organized as follows. The conventional BCH and Zassenhaus formulas are
reviewed in Section~2. In Section~3, after recalling the definition and main properties of
the Jackson $q$-exponential, we determine the most general form of the $q$-Zassenhaus
formula. In Section~4, the explicit form of the operators in the successive factors is
determined up to sixth order to make an appropriate choice for the bases of the
$q$-exponentials. Finally, Section~5 contains the conclusion.\par
%
%
\section{CONVENTIONAL BCH AND ZASSENHAUS FORMULAS}

In the next two sections, we shall make repeated use of the conventional BCH formula for
the product of the exponentials of two (in general) noncommuting operators $X$, $Y$,
\begin{equation}
  \exp(X) \exp(Y) = \exp\left(\sum_{i=1}^{\infty} Z_i\right)  \label{eq:BCH}
\end{equation}
where $Z_i$ is a homogeneous polynomial of degree~$i$ in $X$, $Y$ (therefore said to
be of $i$th order) and
\begin{equation}
  Z_1 = X + Y.
\end{equation}
Since for commuting operators $X$ and $Y$, $Z_1$ is the only nonvanishing term in the
series on the right-hand side of Eq.~(\ref{eq:BCH}), it is obvious that for noncommuting
operators, the additional terms $Z_2$, $Z_3$,~\ldots, all contain the commutator $[X, Y]$.
Following for instance the method given by Wilcox (1967), one can easily find the explicit
expression of the BCH formula up to sixth order:
\begin{eqnarray}
  Z_2 & = & \frac{1}{2} [X, Y] \nonumber \\
  Z_3 & = & \frac{1}{12} \Bigl([X, [X, Y]] - [Y, [X, Y]]\Bigr) \nonumber \\
  Z_4 & = & - \frac{1}{24} [X, [Y, [X, Y]]] \nonumber \\
  Z_5 & = & - \frac{1}{720} [X, [X, [X, [X, Y]]]] - \frac{1}{120} [X, [X, [Y, [X, Y]]]]
       + \frac{1}{360} [Y, [X, [X, [X, Y]]]] \nonumber \\
  && \mbox{} - \frac{1}{360} [X, [Y, [Y, [X, Y]]]] + \frac{1}{120} [Y, [X, [Y, [X, Y]]]]
       + \frac{1}{720} [Y, [Y, [Y, [X, Y]]]]  \nonumber \\
  Z_6 & = & \frac{1}{720} [X, [X, [X, [Y, [X, Y]]]]] - \frac{1}{360} [X, [Y, [X, [X, [X, Y]]]]]
       \nonumber \\
  && \mbox{} + \frac{1}{480} [Y, [X, [X, [X, [X, Y]]]]] - \frac{1}{480} [X, [X, [Y, [Y, [X,
       Y]]]]]  \nonumber \\
  && \mbox{} + \frac{1}{160} [X, [Y, [X, [Y, [X, Y]]]]] - \frac{1}{480} [Y, [X, [X, [Y, [X,
       Y]]]]]  \nonumber \\
  && \mbox{} + \frac{1}{1440} [Y, [Y, [X, [X, [X, Y]]]]] + \frac{1}{288} [X, [Y, [Y, [Y, [X,
       Y]]]]]  \nonumber \\
  && \mbox{} - \frac{1}{180} [Y, [X, [Y, [Y, [X, Y]]]]] + \frac{1}{360} [Y, [Y, [X, [Y, [X,
       Y]]]]].
\end{eqnarray}
\par
%
%
The Zassenhaus formula, which we shall generalize to $q$-exponentials, can be written as
\begin{equation}
  \exp(A+B) = \exp(A) \exp(B) \prod_{i=2}^{\infty} \exp(C_i)  \label{eq:Zas}
\end{equation}
where $C_i$ is a homogeneous polynomial of degree $i$ in $A$, $B$ (therefore said to
be of $i$th order). All the $C_i$'s contain the commutator $[B, A]$ and, using Wilcox
method (Wilcox, 1967) again, they can be determined up to sixth order:
\begin{eqnarray}
  C_2 & = & \frac{1}{2} [B, A] \nonumber \\
  C_3 & = & \frac{1}{3} [[B, A], B] + \frac{1}{6} [[B, A], A] \nonumber \\
  C_4 & = & \frac{1}{8} \Bigl([[[B, A], B], B] + [[[B, A], A], B]\Bigr) + \frac{1}{24}
       [[[B, A], A], A] \nonumber \\
  C_5 & = & \frac{1}{30} \Bigl([[[[B, A], B], B], B] + [[[[B, A], A], A], B]\Bigr) +
       \frac{1}{20} \Bigl([[[[B, A], A], B], B] \nonumber \\
  && \mbox{} + [[[B, A], A], [B, A]]\Bigr) + \frac{1}{10}[[[B, A], B], [B, A]] +
        \frac{1}{120}[[[[B, A], A], A], A] \nonumber \\
  C_6 & = & \frac{1}{144} \Bigl([[[[[B, A], B], B], B], B] + [[[[[B, A], A], A], A], B]\Bigr)
        \nonumber \\
  && \mbox{} + \frac{1}{72} \Bigl([[[[[B, A], A], B], B], B] + [[[[[B, A], A], A], B], B]
        \nonumber \\
  && \mbox{} + [[[[B, A], A], A], [B, A]]\Bigr) + \frac{1}{24} \Bigl([[[[B, A], B], B], 
        [B, A]] \nonumber \\
  && \mbox{} + [[[[B, A], A], B], [B, A]]\Bigr) + \frac{1}{720} [[[[[B, A], A], A], A],
        A].  \label{eq:Zas-C}    
\end{eqnarray}
\par
%
%
It should be noted that the infinite series and products in this section and the next ones
should be understood as formal ones and that their region of convergence should be
studied for any specific choice of operators.\par
%
%
\section{\boldmath GENERAL FORM OF THE $q$-ZASSENHAUS FORMULA}

\setcounter{equation}{0}

The Jackson $q$-exponential $^1$ is defined by (Jackson, 1904)
\begin{equation}
  E_q(z) = \sum_{n=0}^{\infty} \frac{z^n}{[n]_q!}  \label{eq:q-exp} 
\end{equation}
where
\begin{equation}
  [n]_q \equiv \frac{1-q^n}{1-q} = 1 + q + q^2 + \cdots + q^{n-1}
  \label{eq:number}
\end{equation}
and
\begin{equation}
  [n]_q!  \equiv  \left\{\begin{array}{ll}
        1 & {\rm if\ } n=0 \\[0.2cm]
        [n]_q [n-1]_q \ldots [1]_q & {\rm if\ } n=1, 2, \ldots
     \end{array}\right..  
\end{equation}
It has a finite radius of convergence $[\infty]_q = (1-q)^{-1}$ if $0 < q < 1$, but
converges for all finite $z$ if $q > 1$ (Exton, 1983). It is the eigenfunction of the
Jackson $q$-differential operator
\begin{equation}
  D_q E_q(\alpha z) = \alpha E_q(\alpha z)
\end{equation}
where
\begin{equation}
  D_q f(z) \equiv \frac{f(z) - f(qz)}{(1-q)z}
\end{equation}
and it goes over to the conventional exponential for $q \to 1$.\par
%
%
In the appropriate region of definition, the $q$-exponential can be expressed as the
exponential of a series
\begin{equation}
  E_q(z) = \exp\left(\sum_{k=1}^{\infty} c_k(q) z^k\right) \label{eq:series}
\end{equation}
where
\begin{equation}
  c_k(q) = \frac{(1-q)^{k-1}}{k [k]_q}, \qquad k=1, 2, 3, \ldots.  \label{eq:c}
\end{equation}
Although this formula can be traced back to Hardy and Littlewood (1946) and, as quoted
in their paper, may even have been known of other mathematicians before, its simplicity
and usefulness do not seem to have been fully appreciated in the physical literature. For
this reason, in the Appendix, we provide a proof of the formula and derive from it some
other interesting properties of the $q$-exponential.\par
%
%
After these preliminaries, we are now in a position to generalize the Zassenhaus formula
(\ref{eq:Zas}) to the $q$-exponential (\ref{eq:q-exp}) and to obtain the following
result.\par
%
%
{\sl Proposition:} For any choice of $\alpha_i$, $i=0$, 1, 2,~\ldots, such that $\alpha_i
\in \R$, there exists a representation of the $q$-exponential of the sum of two
operators $A$, $B$ as an (in general) infinite product of the form
\begin{equation}
  E_q(A+B) = E_{q^{\alpha_0}}(A) \prod_{i=1}^{\infty} E_{q^{\alpha_i}}(C_i)
  \label{eq:prop} 
\end{equation}
where $C_1 = B$ and $C_i$, $i=2$, 3,~\ldots, are some homogeneous polynomials of
degree $i$ in $A$,~$B$.\par
%
%
{\sl Proof:} Let $\alpha_0$, $\alpha_1$, $\alpha_2$,~\ldots, be any set of real numbers
and define the operator $G^{(0)}$ by
\begin{eqnarray}
  G^{(0)} & \equiv &\left[E_{q^{\alpha_0}}(A)\right]^{-1} E_q(A+B) \nonumber \\
  & = & \exp\left(- \sum_{k=1}^{\infty} c_k\left(q^{\alpha_0}\right) A^k\right)
       \exp\left(\sum_{k=1}^{\infty} c_k(q) (A+B)^k\right)  \label{eq:G0}
\end{eqnarray}
where in the second step we used Eq.~(\ref{eq:series}). The conventional BCH formula
(\ref{eq:BCH}) with $X =- \sum_{k=1}^{\infty} c_k\left(q^{\alpha_0}\right) A^k$ and $Y
= \sum_{k=1}^{\infty} c_k(q) (A+B)^k$ allows one to rewrite $G^{(0)}$ as
\begin{equation}
  G^{(0)} = \exp\left(\sum_{k=1}^{\infty} G^{(0)}_k\right)
\end{equation}
where $G^{(0)}_k$, $k=1$, 2,~\ldots, are some homogeneous polynomials of degree $k$
in $A$, $B$, and $G^{(0)}_1 = B$ since $c_1(q) = c_1\left(q^{\alpha_0}\right) = 1$.\par
%
%
Let now $G^{(1)}$ be defined by
\begin{equation}
  G^{(1)} \equiv \left[E_{q^{\alpha_1}}(C_1)\right]^{-1} G^{(0)}, \qquad 
  C_1 \equiv G^{(0)}_1 = B.  \label{eq:G1}  
\end{equation}
On applying Eqs.~(\ref{eq:series}) and (\ref{eq:BCH}) again, $G^{(1)}$ can be rewritten as
\begin{equation}
  G^{(1)} = \exp\left(\sum_{k=2}^{\infty} G^{(1)}_k\right)  \label{eq:G1bis}
\end{equation}
where $G^{(1)}_k$, $k=2$, 3,~\ldots, are some homogeneous polynomials of degree $k$
in $A$, $B$, and there is no first-order term due to the choice made for $C_1$.
Equations (\ref{eq:G0}) and (\ref{eq:G1}) together lead to the relation
\begin{equation}
  E_q(A+B) = E_{q^{\alpha_0}}(A) E_{q^{\alpha_1}}(C_1) G^{(1)}  \label{eq:Eq1}
\end{equation}
where $G^{(1)}$ is given in (\ref{eq:G1bis}).\par
%
%
Let us then assume that for some $j \in \R^+$, the relations
\begin{equation}
  E_q(A+B) = E_{q^{\alpha_0}}(A) \left(\prod_{i=1}^j E_{q^{\alpha_i}}(C_i)\right) G^{(j)}  
  \label{eq:Eqj}
\end{equation}
\begin{equation}
  G^{(j)} = \exp\left(\sum_{k=j+1}^{\infty} G^{(j)}_k\right)  \label{eq:Gj}
\end{equation}
hold for some homogeneous polynomials $C_i$ (resp.~$G^{(j)}_k$) of degree $i$
(resp.~k) in $A$, $B$. On setting
\begin{equation}
  G^{(j+1)} \equiv \left[E_{q^{\alpha_{j+1}}}(C_{j+1})\right]^{-1} G^{(j)}, \qquad 
  C_{j+1} \equiv G^{(j)}_{j+1} 
\end{equation}
and using Eqs.~(\ref{eq:series}) and (\ref{eq:BCH}), we get
\begin{equation}
  G^{(j+1)} = \exp\left(\sum_{k=j+2}^{\infty} G^{(j+1)}_k\right)  
\end{equation}
where $G^{(j+1)}_k$ are some homogeneous polynomials of degree $k$ in $A$, $B$.
Hence Eqs.~(\ref{eq:Eqj}) and (\ref{eq:Gj}) are valid when $j$ is replaced by $j+1$.
Furthermore, as shown in (\ref{eq:G1bis}) and (\ref{eq:Eq1}), they hold for $j=1$. This
therefore completes their proof by induction over $j$.\par
%
%
{}For $j \to \infty$, we finally get the representation (\ref{eq:prop}) of $E_q(A+B)$ as a
formal infinite product.\par
%
%
\section{\boldmath EXPLICIT FORM OF THE $q$-ZASSENHAUS FORMULA UP TO SIXTH
ORDER}

\setcounter{equation}{0}

In the present section, we will apply the method presented in the previous one to
determine the explicit form of the first few operators $C_i$. We will then make a
choice for the bases $q^{\alpha_i}$ in order to get the simplest formula.\par
%
%
To start with, since the $q$-deformed counterpart of the multiplicative property of the
conventional exponential reads (Sch\"utzenberger, 1953; Cigler, 1979; Fairlie and Wu,
1997) 
\begin{equation}
  E_q(A) E_q(B) = E_q(A+B) \qquad \mbox{\rm if}\quad [B,A]_q \equiv BA - qAB = 0
\end{equation}
it is convenient to choose $\alpha_0 = \alpha_1 = 1$ in Eq.~(\ref{eq:prop}). In this way,
all the operators $C_i$, $i=2$, 3,~\ldots, will contain the $q$-commutator $[B, A]_q$
and therefore no terms depending only on $A$ or $B$.\par
%
%
{}For $j=0$, 1, \ldots, $n-1$ successively, we then determine the polynomials
$G^{(j)}_k$, $k=j+1$, $j+2$, \ldots,~$n$, of Eq.~(\ref{eq:Gj}) up to some maximal order
$n$. This provides us with some explicit expressions for $C_i = G^{(i-1)}_i$, $i=2$, 3,
\ldots,~$n$, in terms of $A$, $B$ and of the coefficients $c_k(q)$,
$c_k\left(q^{\alpha_2}\right)$,~\ldots, defined in Eq.~(\ref{eq:c}). Taking the latter
into account, the $C_i$'s are finally re-expressed in terms of $q$-commutators.\par
%
%
Let us illustrate the procedure by giving detailed results for $n=3$. We successively get
\begin{eqnarray}
  G^{(0)}_1 & = & B  \nonumber \\
  G^{(0)}_2 & = & c_2(q) B^2 + \left[c_2(q) + \frac{1}{2}\right] BA + \left[c_2(q) -
        \frac{1}{2}\right] AB   \nonumber \\
  G^{(0)}_3 & = & c_3(q) B^3 + \left[c_3(q) + \frac{1}{2} c_2(q) - \frac{1}{12}\right] 
        B^2 A + \left[c_3(q) + \frac{1}{6}\right] BAB \nonumber \\
  && \mbox{} + \left[c_3(q) - \frac{1}{2} c_2(q) - \frac{1}{12}\right] AB^2 + \left[c_3(q)
        + c_2(q) + \frac{1}{6}\right] BA^2 + \left[c_3(q) - \frac{1}{3}\right] ABA 
        \nonumber \\
  && \mbox{} + \left[c_3(q) - c_2(q) + \frac{1}{6}\right] A^2 B  \nonumber \\
  G^{(1)}_2 & = & \left[c_2(q) + \frac{1}{2}\right] BA + \left[c_2(q) -
        \frac{1}{2}\right] AB   \nonumber \\
  G^{(1)}_3 & = & \left[c_3(q) - \frac{1}{3}\right] B^2 A + \left[c_3(q) +
        \frac{2}{3}\right] BAB + \left[c_3(q) - \frac{1}{3}\right] AB^2 \nonumber \\
  && \mbox{} + \left[c_3(q) + c_2(q) + \frac{1}{6}\right] BA^2 + \left[c_3(q) -
        \frac{1}{3}\right] ABA + \left[c_3(q) - c_2(q) + \frac{1}{6}\right] A^2 B
        \nonumber \\
  G^{(2)}_3 & = & G^{(1)}_3
\end{eqnarray}
from which we deduce that
\begin{equation}
  C_2 = G^{(1)}_2 = \frac{1}{[2]_q} (BA - qAB) = \frac{1}{[2]_q} [B, A]_q
  \label{eq:C2} 
\end{equation}
and
\begin{eqnarray}
  C_3 & = & G^{(2)}_3 \nonumber \\
  & = & \frac{1}{[3]_q} \left[-q B^2 A + (1+q^2) BAB - q AB^2\right] + \frac{1}{[3]_q!}
        \left[BA^2 - q (1+q) ABA + q^3 A^2 B \right] \nonumber \\
  & = & \frac{1}{[3]_q} [[B, A]_q, B]_q + \frac{1}{[3]_q!} [B, A]_q, A]_{q^2}.  
\end{eqnarray}
Such expressions, which do not depend on $\alpha_2$, $\alpha_3$,~\ldots, coincide with
those given by Katriel {\sl et al.} (1996) and by Sridhar and Jagannathan (2002).\par
%
%
With the help of Mathematica, we have calculated in the same way the next terms up to
sixth order. They can be written as
\begin{eqnarray}
  C_4 & = & \frac{1}{[2]_q [4]_q} \Bigl([[[B, A]_q, B]_q, B]_{q^2} + [[[B, A]_q, A]_{q^2},
        B]_q \Bigr) \nonumber \\
  && + \frac{1}{[4]_q!} [[[B, A]_q, A]_{q^2}, A]_{q^3} + \frac{q^a}{[2]_q [4]_q [2]_{q^a}}
        [[B, A]_q, [B, A]_q]_{q^{2-a}}  \label{eq:C4}     
\end{eqnarray}
\begin{eqnarray}
  C_5 & = & \frac{1}{[3]_q!\, [5]_q} \Bigl([[[[B, A]_q, B]_q, B]_{q^2}, B]_{q^3} + [[[[B, A]_q,
        A]_{q^2}, A]_{q^3}, B]_q \Bigr) \nonumber \\
  && + \frac{1}{[2]_q^2\, [5]_q} \Bigl([[[[B, A]_q, A]_{q^2}, B]_q, B]_{q^2} + [[[B, A]_q,
        A]_{q^2}, [B, A]_q]_{q^2} \Bigr) \nonumber \\ 
  && + \frac{1}{[2]_q [5]_q} [[[B, A]_q, B]_q, [B, A]_q]_{q^2} +  \frac{1}{[5]_q!} [[[[B, A]_q,
        A]_{q^2}, A]_{q^3}, A]_{q^4}  
\end{eqnarray}
\begin{eqnarray}
  C_6 & = & \frac{1}{[4]_q!\, [6]_q} \Bigl([[[[[B, A]_q, B]_q, B]_{q^2}, B]_{q^3}, B]_{q^4} +
        [[[[[B, A]_q, A]_{q^2}, A]_{q^3}, A]_{q^4}, B]_q \Bigr) \nonumber \\
  && + \frac{1}{[2]_q^2\, [3]_q [6]_q} \Bigl([[[[[B, A]_q, A]_{q^2}, B]_q, B]_{q^2}, B]_{q^3} +
        [[[[[B, A]_q, A]_{q^2}, A]_{q^3}, B]_q, B]_{q^2} \nonumber \\ 
  && + [[[[B, A]_q, A]_{q^2}, A]_{q^3}, [B, A]_q]_{q^2} \Bigr) + \frac{1}{[2]_q^2\, [6]_q} 
        \Bigl([[[[B, A]_q, B]_q, B]_{q^2}, [B, A]_q]_{q^2} \nonumber \\
  && + [[[[B, A]_q, A]_{q^2}, B]_q, [B, A]_q]_{q^2}\Bigr) + \frac{1}{[6]_q!} [[[[[B, A]_q,
        A]_{q^2}, A]_{q^3}, A]_{q^4}, A]_{q^5} \nonumber \\
  && + \frac{q^a}{[2]_q^2\, [6]_q [3]_{q^a}} [[[B, A]_q, [B, A]_q]_{q^{2-a}}, [B,
        A]_q]_{q^{2+a}}  \nonumber \\
  && + \frac{q^b}{[3]_q [6]_q [2]_{q^b}} [[[B, A]_q, B]_q, [[B, A]_q, B]_q]_{q^{3-b}} 
        \nonumber \\
  && + \frac{q^b}{[3]_q!\, [6]_q [2]_{q^b}} \Bigl([[[B, A]_q, B]_q, [[B, A]_q,
        A]_{q^2}]_{q^{3-b}} + [[[B, A]_q, A]_{q^2}, [[B, A]_q, B]_q]_{q^{3-b}}\Bigr)  
        \nonumber \\
  && + \frac{q^b}{[2]_q^2\,[3]_q [6]_q [2]_{q^b}} [[[B, A]_q, A]_{q^2}, [[B, A]_q,
        A]_{q^2}]_{q^{3-b}}  \label{eq:C6}       
\end{eqnarray}
where we have set $a \equiv \alpha_2$, $b\equiv \alpha_3$. It can be easily checked that
for $q \to 1$, Eqs.~(\ref{eq:C2}) -- (\ref{eq:C6}) give back the conventional results given in
Eq.~(\ref{eq:Zas-C}).\par
%
%
{}From such general results, it is clear that the simplest forms for $C_4$ and $C_6$
correspond to the choice $a=2$, $b=3$, in which case they become
\begin{equation}
  C_4 = \frac{1}{[2]_q [4]_q} \Bigl([[[B, A]_q, B]_q, B]_{q^2} + [[[B, A]_q, A]_{q^2},
        B]_q \Bigr) + \frac{1}{[4]_q!} [[[B, A]_q, A]_{q^2}, A]_{q^3}  \label{eq:C4-SJ}
\end{equation}
\begin{eqnarray}
  C_6 & = & \frac{1}{[4]_q!\, [6]_q} \Bigl([[[[[B, A]_q, B]_q, B]_{q^2}, B]_{q^3}, B]_{q^4} +
        [[[[[B, A]_q, A]_{q^2}, A]_{q^3}, A]_{q^4}, B]_q \Bigr) \nonumber \\
  && + \frac{1}{[2]_q^2\, [3]_q [6]_q} \Bigl([[[[[B, A]_q, A]_{q^2}, B]_q, B]_{q^2}, B]_{q^3} +
        [[[[[B, A]_q, A]_{q^2}, A]_{q^3}, B]_q, B]_{q^2} \nonumber \\ 
  && + [[[[B, A]_q, A]_{q^2}, A]_{q^3}, [B, A]_q]_{q^2} \Bigr) + \frac{1}{[2]_q^2\, [6]_q} 
        \Bigl([[[[B, A]_q, B]_q, B]_{q^2}, [B, A]_q]_{q^2} \nonumber \\
  && + [[[[B, A]_q, A]_{q^2}, B]_q, [B, A]_q]_{q^2}\Bigr) + \frac{1}{[6]_q!} [[[[[B, A]_q,
        A]_{q^2}, A]_{q^3}, A]_{q^4}, A]_{q^5} 
\end{eqnarray}
while $C_5$ is independent of the choice made for the bases. Equation (\ref{eq:C4-SJ})
agrees with Sridhar and Jagannathan (2002), who stopped at fourth order. The calculation of
the next two terms, which we have carried out in the present paper, strengthens the
conjecture made by these authors according to which $\alpha_i$, $i=2$, 3,~\ldots, should
be taken as $\alpha_i = i$.\par
%
%
In contrast, it is clear from Eqs.~(\ref{eq:C4}) and (\ref{eq:C6}) that the choice $\alpha_i =
1$, $i=2$, 3,~\ldots, made by Katriel {\sl et al.} (1996) leads to much more complicated
expressions for $C_4$ and $C_6$, given by
\begin{eqnarray}
  C_4 & = & \frac{1}{[2]_q [4]_q} \Bigl([[[B, A]_q, B]_q, B]_{q^2} + [[[B, A]_q, A]_{q^2},
        B]_q \Bigr) \nonumber \\
  && + \frac{1}{[4]_q!} [[[B, A]_q, A]_{q^2}, A]_{q^3} + \frac{q}{[2]_q^2\, [4]_q}
        [[B, A]_q, [B, A]_q]_q  \label{eq:C4-K}       
\end{eqnarray}
\begin{eqnarray}
  C_6 & = & \frac{1}{[4]_q!\, [6]_q} \Bigl([[[[[B, A]_q, B]_q, B]_{q^2}, B]_{q^3}, B]_{q^4} +
        [[[[[B, A]_q, A]_{q^2}, A]_{q^3}, A]_{q^4}, B]_q \Bigr) \nonumber \\
  && + \frac{1}{[2]_q^2\, [3]_q [6]_q} \Bigl([[[[[B, A]_q, A]_{q^2}, B]_q, B]_{q^2}, B]_{q^3} +
        [[[[[B, A]_q, A]_{q^2}, A]_{q^3}, B]_q, B]_{q^2} \nonumber \\ 
  && + [[[[B, A]_q, A]_{q^2}, A]_{q^3}, [B, A]_q]_{q^2} \Bigr) + q 
        [[[B, A]_q, [B, A]_q]_q, [B, A]_q]_{q^3} \nonumber \\
  && + q [[[B, A]_q, B]_q, [[B, A]_q, A]_{q^2}]_{q^2} + q [[[B, A]_q, A]_{q^2}, [[B, A]_q,
        B]_q]_{q^2}\Bigr)  \nonumber \\
  && + \frac{1}{[2]_q^2\, [6]_q} \Bigl([[[[B, A]_q, B]_q, B]_{q^2}, [B, A]_q]_{q^2}
        + [[[[B, A]_q, A]_{q^2}, B]_q, [B, A]_q]_{q^2}\Bigr)  \nonumber \\
  && + \frac{1}{[6]_q!} [[[[[B, A]_q, A]_{q^2}, A]_{q^3}, A]_{q^4}, A]_{q^5} + 
       \frac{q}{[3]_q!\, [6]_q} [[[B, A]_q, B]_q, [[B, A]_q, B]_q]_{q^2} \nonumber \\
  && + \frac{q}{[2]_q^3\,[3]_q [6]_q} [[[B, A]_q, A]_{q^2}, [[B, A]_q,
        A]_{q^2}]_{q^2}         
\end{eqnarray}
respectively. Equation (\ref{eq:C4-K}) actually corrects their final expression, where we have
found a misprint in one of the terms.\par
%
%
\section{CONCLUSION}

In the present paper, we have revisited the $q$-deformed counterpart of the Zassenhaus
formula that arises on replacing the conventional exponential of the sum of two
noncommuting operators $A$ and $B$ by the Jackson $q$-exponential of the sum of two
non-$q$-commuting ones.\par
%
%
We have proved that for any choice $q^{\alpha_i}$, $\alpha_i \in \R$, $i=0$, 1, 2,~\ldots,
of the bases of the $q$-exponentials, there exists a representation of $E_q(A+B)$ as an (in
general) infinite product $E_{q^{\alpha_0}}(A) E_{q^{\alpha_1}}(B) \prod_{i=2}^{\infty}
E_{q^{\alpha_i}}(C_i)$, where $C_i$ are some operators of increasing order in $A$,
$B$.\par
%
%
To reproduce the multiplicative property of the $q$-exponential for $q$-commutative
operators we have then selected $\alpha_0 = \alpha_1 = 1$. With this choice and leaving
the remaining $\alpha_i$'s arbitrary, we have finally obtained the explicit form of the
$C_i$'s for $i=2$, 3,~\ldots, 6. This has unambigously shown that $\alpha_2 = 2$ and
$\alpha_3 = 3$ lead to the simplest $q$-Zassenhaus formula up to sixth order. Our work
therefore confirms and reinforces the Sridhar and Jagannathan (2002) conjecture, derived
from a study of fourth-order terms, according to which the best choice should be $\alpha_i
= i$, $i=2$, 3,~\ldots.\par
%
%
\section*{ACKNOWLEDGMENTS}

The author is indebted to R.\ Jagannathan, K.A.\ Penson, T.\ Prellberg, and K.\ Srinivasa Rao
for some interesting comments and discussions. She is a Research Director of the National
Fund for Scientific Research (FNRS), Belgium.\par
%
%
\section*{APPENDIX}

\renewcommand{\theequation}{A.\arabic{equation}}
\setcounter{equation}{0}

In this appendix we demonstrate Eqs.~(\ref{eq:series}) and (\ref{eq:c}), then use them
to obtain some interesting properties of the $q$-exponential.\par
%
%
Proving Eqs.~(\ref{eq:series}) and (\ref{eq:c}) amounts to finding the Taylor expansion of
the logarithm of the $q$-exponential
\begin{equation}
  \ln E_q(z) = \sum_{k=1}^{\infty} c_k(q) z^k. 
\end{equation}
\par
%
%
In Pourahmadi (1986) (see also Sachkov, 1996), it has been shown that if the functions $f(z)
= \sum_{k=0}^{\infty} a_k z^k$ and $h(z) = \ln f(z) = \sum_{k=1}^{\infty} c_k z^k$,
where $a_0=1$, are analytic in some neighbourhood of zero, then the Taylor coefficients
of $h(z)$ satisfy the recursion relation
\begin{equation}
  c_k = a_k - \frac{1}{k} \sum_{j=1}^{k-1} j a_{k-j} c_j, \qquad k=2, 3, \ldots
\end{equation}
with $c_1 = a_1$. Applying this result to the logarithm of the $q$-exponential leads to
the relations
\begin{eqnarray}
  c_k(q) & = & \frac{1}{[k]_q!} - \frac{1}{k} \sum_{j=1}^{k-1} \frac{j}{[k-j]_q!} c_j(q),
        \qquad k=2, 3, \ldots \label{eq:recursion}\\
  c_1(q) & = & 1. \label{eq:condition}
\end{eqnarray}
\par
%
%
It is straightforward to check that the solution of Eq.~(\ref{eq:recursion}), satisfying
condition (\ref{eq:condition}), is provided by Eq.~(\ref{eq:c}). Inserting such an expression in
Eq.~(\ref{eq:recursion}) converts the latter into the relation
\begin{equation}
  \sum_{j=1}^k  
     \left[\begin{array}{c}
         k \\ j
     \end{array}\right]_q
  (1-q)^{j-1} [j-1]_q! = k, \qquad k=2, 3, \ldots \label{eq:relation}
\end{equation}
where
\begin{equation}
  \left[\begin{array}{c}        
      k \\ j
  \end{array}\right]_q \equiv \frac{[k]_q!}{[j]_q!\, [k-j]_q!}
\end{equation}
denotes a $q$-binomial coefficient (Exton, 1983). Equation (\ref{eq:relation}) can be
easily proved by induction over $k$ by using the recursion relation
\begin{equation}
  \left[\begin{array}{c}        
      k \\ j
  \end{array}\right]_q 
  = q^j   \left[\begin{array}{c}        
                 k-1 \\ j
             \end{array}\right]_q
  + \left[\begin{array}{c}        
         k-1 \\ j-1
     \end{array}\right]_q, \qquad j=1, 2, \ldots, k-1.
\end{equation}
We indeed obtain
\begin{eqnarray}
  \lefteqn{\sum_{j=1}^k  
       \left[\begin{array}{c}                              
           k \\ j
       \end{array}\right]_q (1-q)^{j-1} [j-1]_q!} \nonumber\\
  & = & \sum_{j=1}^{k-1}  
       \left[\begin{array}{c}                              
           k-1 \\ j
       \end{array}\right]_q q^j (1-q)^{j-1} [j-1]_q!
       + \sum_{j=0}^{k-2}  
       \left[\begin{array}{c}                              
           k-1 \\ j
       \end{array}\right]_q (1-q)^j [j]_q! \nonumber \\
  && \mbox{} + (1-q)^{k-1} [k-1]_q! \nonumber \\
  & = & \sum_{j=1}^{k-1}  
       \left[\begin{array}{c}                              
           k-1 \\ j
       \end{array}\right]_q q^j (1-q)^{j-1} [j-1]_q!
       + 1 \nonumber \\
  && \mbox{} + \sum_{j=1}^{k-1}  
       \left[\begin{array}{c}                              
           k-1 \\ j
       \end{array}\right]_q (1-q)^{j-1} (1-q^j) [j-1]_q! \nonumber \\
  & = & \sum_{j=1}^{k-1}  
       \left[\begin{array}{c}                              
           k-1 \\ j
       \end{array}\right]_q (1-q)^{j-1} [j-1]_q! + 1 \nonumber \\
  & = & (k-1) + 1
\end{eqnarray}
where in the last step use has been made of the induction hypothesis.\par
%
%
Equations (\ref{eq:series}) and (\ref{eq:c}) can be applied to derive the following properties of
the $q$-exponential, already quoted in (Ubriaco, 1992) (where we have corrected some
misprints),
\begin{equation}
  E_q(z) E_{q^{-1}}(z) = 1  \label{eq:property1}
\end{equation}
\begin{equation}
  E_q(z) E_q(-z) = E_{q^2}\left(\frac{1-q}{1+q}z^2\right)  \label{eq:property2}
\end{equation}
\begin{equation}
  E_q([n]_q z) = \prod_{m=0}^{n-1} E_{q^n}(q^m z), \qquad n=2, 3, \ldots
  \label{eq:property3}
\end{equation}
as well as a generalization of Eq.~(\ref{eq:property2}),
\begin{equation}
  \prod_{m=0}^{n-1} E_q(e^{2\pi {\rm i}m/n} z) = E_{q^n}\left(\frac{(1-q)^{n-1}}
  {[n]_q} z^n\right), \qquad n=2, 3, \ldots.
\end{equation}
The proof of these relations is based upon the multiplicative property of the
ordinary exponential, $\exp(x) \exp(y) = \exp(x+y)$, and on some elementary properties of the
coefficients $c_k(q)$, defined in (\ref{eq:c}),
\begin{equation}
  c_k(q^{-1}) = (-1)^{k-1} c_k(q)
\end{equation}
\begin{equation}
  2 c_{2k}(q) = \left(\frac{1-q}{1+q}\right)^k c_k(q^2)
\end{equation}
\begin{equation}
  [n]_{q^k} c_k(q^n) = ([n]_q)^k c_k(q)
\end{equation}
\begin{equation}
  n c_{nk}(q) = \left(\frac{(1-q)^{n-1}}{[n]_q}\right)^k c_k(q^n).
\end{equation}
\par
%
%
\newpage
\section*{REFERENCES}
\setlength{\parindent}{0cm}

Baker, H.\ F.\ (1902). {\em Proc.\ London Math.\ Soc.}, {\bf 34}, 347.

Baker, H.\ F.\ (1903). {\em Proc.\ London Math.\ Soc.}, {\bf 35}, 333.

Baker, H.\ F.\ (1904a). {\em Proc.\ London Math.\ Soc.}, {\bf 2}, 293.

Baker, H.\ F.\ (1904b). {\em Proc.\ London Math.\ Soc.}, {\bf 3}, 24.

Brif, C.\ (1996). {\em Phys.\ Rev.\ A}, {\bf 54}, 5253.

Campbell, J.\ E.\ (1898). {\em Proc.\ London Math.\ Soc.}, {\bf 29}, 14.

Chaichian, M., and Demichev, A.\ (1996). {\em Introduction to Quantum Groups}, World
Scientific, Singapore.

Cigler, J.\ (1979). {\em Monatsch.\ Math.}, {\bf 88}, 87.

Drinfeld, V.\ G.\ (1987). In {\em Proceedings of the International Congress of
Mathematicians, Berkeley, 1986}, A.\ M.\ Gleason, ed., American Mathematical Society,
Providence, RI, p.\ 798.

Exton, H.\ (1983). {\em $q$-Hypergeometric Functions and Applications}, Ellis Horwood,
Chichester.

Faddeev, L., Reshetikhin, N., and Takhtajan, L.\ (1988). In {\em Algebraic Analysis}, M.\
Kashiwara and T.\ Kawai, eds., Academic Press, New York, Vol.\ 1, p.\ 129.

Fairlie, D., and Wu, M.\ (1997). The reversed $q$-exponential functional relation,
q-alg/9704013.

Hardy, G.\ H., and Littlewood, J.\ E.\ (1946). {\em Proc.\ Cambridge Phil.\ Soc.}, {\bf 42}, 85.

Hatano, N., and Suzuki, M.\ (1991). {\em Phys.\ Lett.\ A}, {\bf 153}, 191.

Hausdorff, F.\ (1906). {\em Ber.\ Verhandl.\ Saechs.\ Akad.\ Wiss.\ Leipzig, Math.-Naturw.\
Kl.}, {\bf 58}, 19.

Heine, E.\ (1847). {\em J.\ Reine Angew.\ Math.}, {\bf 34}, 285.

Jackson, F.\ H.\ (1904). {\em Edin.\ Math.\ Soc.}, {\bf 22}, 28.

Jimbo, M.\ (1985). {\em Lett.\ Math.\ Phys.}, {\bf 10}, 63. 

Jimbo, M.\ (1986). {\em Lett.\ Math.\ Phys.}, {\bf 11}, 247. 

Katriel, J., Rasetti, M., and Solomon, A.\ I.\ (1996). {\em Lett.\ Math.\ Phys.}, {\bf 37}, 11.

Katriel, J., and Solomon, A.\ I.\ (1991). {\em J.\ Phys.\ A}, {\bf 24}, L1139.

Klimyk, A., and Schm\"udgen, K.\ (1997). {\em Quantum Groups and Their Representations},
Springer, Berlin.

Magnus, W.\ (1954). {\em Comm.\ Pure Appl.\ Math.}, {\bf 7}, 649.

Majid, S.\ (1995). {\em Foundations of Quantum Group Theory}, Cambridge University Press,
Cambridge.

Pourahmadi, M.\ (1984). {\em Amer.\ Math.\ Monthly}, {\bf 91}, 303.

Sachkov, V.\ N.\ (1996). {\em Combinatorial Methods in Discrete Mathematics}, Cambridge
University Press, Cambridge.

Sch\"utzenberger, M.\ P.\ (1953). {\em C.\ R.\ Acad.\ Sci.\ Paris}, {\bf 236}, 352.

Sridhar, R., and Jagannathan, R.\ (2002). On the $q$-analogues of the Zassenhaus formula for
disentangling exponential operators, math-ph/0212068, to be published in {\em Proceedings
of the International Conference on Special Functions, Chennai, 2002}. 

Suzuki, M.\ (1977). {\em Commun.\ Math.\ Phys.}, {\bf 57}, 193.

Ubriaco, M.\ R.\ (1992). {\em Phys.\ Lett.\ A}, {\bf 163}, 1.

Wilcox, R.\ M.\ (1967). {\em J.\ Math.\ Phys.}, {\bf 8}, 962.

Witschel, W.\ (1975). {\em J.\ Phys.\ A}, {\bf 8}, 143.

Zhao, Z.\ S.\ (1991). {\em J.\ Math.\ Phys.}, {\bf 32}, 2783.
%
%
\newpage
\section*{Footnote}

$^1$ Jackson has actually introduced two different kinds of $q$-exponentials, related to one
another by inversion. The function considered in Eq.~(\ref{eq:q-exp}) is connected to one of
them. It is referred to as the Jackson $q$-exponential in most works of modern physics. 

\end{document}